\newcommand{\la}{\langle}
\newcommand{\ra}{\rangle}

\newcommand{\beq}{\begin{eqnarray}}
\newcommand{\eeq}{\end{eqnarray}}

\renewcommand{\theequation}{\thesection.\arabic{equation}}
\newcommand{\do}{^{\circ }}
\newcommand{\cl}{\centerline}

\newcommand{\btem}{\bibitem}

\newcommand{\TK}{T.\ Kunihiro}
\newcommand{\PL}{Phys.\ Lett.\ {\bf B}}
\newcommand{\PTP}{Prog.\ Theor.\ Phys.}

\newcommand{\PRL}{Phys.\ Rev. \ Lett.}

\newcommand{\HK}{T. Hatsuda and T. Kunihiro}

\documentstyle[12pt]{article}

\catcode`\@=11
\long\def\@makefntext#1{
\protect\noindent \hbox to 3.2pt {\hskip-.9pt
$^{{\ninerm\@thefnmark}}$\hfil}#1\hfill}		

\def\@makefnmark{\hbox to 0pt{$^{\@thefnmark}$\hss}}  

\def\ps@myheadings{\let\@mkboth\@gobbletwo
\def\@oddhead{\hbox{}
\rightmark\hfil\ninerm\thepage}
\def\@oddfoot{}\def\@evenhead{\ninerm\thepage\hfil
\leftmark\hbox{}}\def\@evenfoot{}
\def\sectionmark##1{}\def\subsectionmark##1{}}

\renewcommand{\thefootnote}{\fnsymbol{footnote}}

\newcounter{sectionc}\newcounter{subsectionc}\newcounter{subsubsectionc}
\renewcommand{\section}[1] {\vspace*{0.6cm}\addtocounter{sectionc}{1}
\setcounter{subsectionc}{0}\setcounter{subsubsectionc}{0}\noindent
	{\normalsize\bf\thesectionc. #1}\par\vspace*{0.4cm}}
\renewcommand{\subsection}[1] {\vspace*{0.6cm}\addtocounter{subsectionc}{1}
	\setcounter{subsubsectionc}{0}\noindent
	{\normalsize\it\thesectionc.\thesubsectionc. #1}\par\vspace*{0.4cm}}
\renewcommand{\subsubsection}[1]
{\vspace*{0.6cm}\addtocounter{subsubsectionc}{1}
	\noindent {\normalsize\rm\thesectionc.\thesubsectionc.
 \thesubsubsectionc.
	#1}\par\vspace*{0.4cm}}

\newcounter{appendixc}
\newcounter{subappendixc}[appendixc]
\newcounter{subsubappendixc}[subappendixc]

\renewcommand{\appendix}[1] {\vspace*{0.6cm}
        \refstepcounter{appendixc}
        \setcounter{figure}{0}
        \setcounter{table}{0}
        \setcounter{equation}{0}
        \renewcommand{\thefigure}{\Alph{appendixc}.\arabic{figure}}
        \renewcommand{\thetable}{\Alph{appendixc}.\arabic{table}}
        \renewcommand{\theappendixc}{\Alph{appendixc}}
        \renewcommand{\theequation}{\Alph{appendixc}.\arabic{equation}}
        \noindent{\bf Appendix \theappendixc #1}\par\vspace*{0.4cm}}

\def\abstracts#1{{
	\centering{\begin{minipage}{12.2truecm}\footnotesize\baselineskip=
      12pt\noindent
	\centerline{\footnotesize ABSTRACT}\vspace*{0.3cm}
	\parindent=0pt #1
	\end{minipage}}\par}}

\renewenvironment{thebibliography}[1]
	{\begin{list}{\arabic{enumi}.}
	{\usecounter{enumi}\setlength{\parsep}{0pt}
\setlength{\leftmargin 1.25cm}{\rightmargin 0pt}
	 \setlength{\itemsep}{0pt} \settowidth
	{\labelwidth}{#1.}\sloppy}}{\end{list}}

\newcounter{itemlistc}
\newcounter{romanlistc}
\newcounter{alphlistc}
\newcounter{arabiclistc}

\newcommand{\fcaption}[1]{
        \refstepcounter{figure}
        \setbox\@tempboxa = \hbox{\footnotesize Fig.~\thefigure. #1}
        \ifdim \wd\@tempboxa > 6in
           {\begin{center}
        \parbox{6in}{\footnotesize\baselineskip=12pt Fig.~\thefigure. #1}
            \end{center}}
        \else
             {\begin{center}
             {\footnotesize Fig.~\thefigure. #1}
              \end{center}}
        \fi}

\newcommand{\tcaption}[1]{
        \refstepcounter{table}
        \setbox\@tempboxa = \hbox{\footnotesize Table~\thetable. #1}
        \ifdim \wd\@tempboxa > 6in
           {\begin{center}
        \parbox{6in}{\footnotesize\baselineskip=12pt Table~\thetable. #1}
            \end{center}}
        \else
             {\begin{center}
             {\footnotesize Table~\thetable. #1}
              \end{center}}
        \fi}

\def\@citex[#1]#2{\if@filesw\immediate\write\@auxout
	{\string\citation{#2}}\fi
\def\@citea{}\@cite{\@for\@citeb:=#2\do
	{\@citea\def\@citea{,}\@ifundefined
	{b@\@citeb}{{\bf ?}\@warning
	{Citation `\@citeb' on page \thepage \space undefined}}
	{\csname b@\@citeb\endcsname}}}{#1}}

\newif\if@cghi
\def\cite{\@cghitrue\@ifnextchar [{\@tempswatrue
	\@citex}{\@tempswafalse\@citex[]}}
\def\citelow{\@cghifalse\@ifnextchar [{\@tempswatrue
	\@citex}{\@tempswafalse\@citex[]}}
\def\@cite#1#2{{$\null^{#1}$\if@tempswa\typeout
	{IJCGA warning: optional citation argument
	ignored: `#2'} \fi}}

 1
 1
 1

\font\ninerm=cmr9

\begin{document}
%
%
%
\centerline{\normalsize\bf Possible  Production of the Sigma Meson in Nuclei
\footnote{The major part of this report is based on 
 \cite{rcnp94} and a part of \cite{suppl}.}}

\centerline{\footnotesize TEIJI KUNIHIRO}
\baselineskip=13pt
\centerline{\footnotesize\it  
Faculty of Science and Technology, Ryukoku University, Seta
 Ohtsu, 520-21, Japan}

\vspace*{0.9cm}
\abstracts{It should be a typical subject in this workshop to explore possible
 change of hadron  properties in association with that of QCD vacuum due to an 
environmental change.
 In this report, we first summarize what hadrons are expected to show such a
 change and how they do. Then 
we propose several experiments including the ones utilizing electron and 
 photon beams
 for examining possible restoration of chiral
 symmetry in nuclei and the possible existence of the sigma meson.
The experiments are based on the observation that the sigma meson may
 decrease  the mass and the width along with the partial restoration of chiral
symmetry in the medium.}
\normalsize\baselineskip=15pt
\setcounter{footnote}{0}
\renewcommand{\thefootnote}{\alph{footnote}}

\section{Introduction}
    
 An underlying observation for this workshop on 
``Quark Nuclear  Physics'' can  be
 the following\cite{physrep}: (1) The QCD vacuum will change under an
 environmental change, 
 characterized by the baryonic density $\rho_B$, temperature
  $T$, external gauge fields such  as magnetic field $\vec {B}$.
 (2) According to a general principle of the quantum field theory, 
 a  change of the vacuum may 
 reflect in those of the elementary excitations, hadrons for the QCD vacuum.
A nucleus is  of course  a typical system with  $\rho_B$.
  
The QCD vacuum  is  characterized by  the confinement of the colored
 objects, dynamical breaking of chiral symmetry, $U_A(1)$ anomaly and so on.
 The rules extracted from hadron phenomenology such as the vector-meson
dominance (VMD) and the Okubo-Zweig-Iizuka (OZI) rule might be also related 
with some fundamental properties of the QCD vacuum.

The problems are  then what hadrons  change the properties, and 
 how they do in hot and/or dense medium.
  One should also ask how  they are detected in experiment.
Such a theoretical study was  initiated about a decade ago 
by several authors  independently\cite{pis,pl84}; see \cite{physrep}
 for a review. 
 In Table 1, we list up several hadrons and their expected 
 behaviors as the fundamental properties of the QCD vacuum 
change\cite{physrep}.
In association with (partial) restoration of chiral symmetry, the mass of
 the $\sigma$ meson\cite{pl87,bernard}
 is expected to decrease.  
 Some people\cite{rho} expect that the vector mesons $\rho, \omega$ and $\phi$
 also show a decrease of their masses 
in association with the chiral restoration.\footnote{
However see also \cite{yabu}.
 The vector meson dominance (VMD) principle might be  or might not be
  modified at $T\not=0$ and/or $\rho\not=0$.\cite{yabu} This is related
 with the behavior of the vector mesons in hot and/or dense medium.} 
 \ The $U_A(1)$ anomaly, which is responsible for 
 lifting the $\eta '$ meson mass as high as about 1 GeV and make the 
 $\eta' (\eta)$ almost flavor singlet (octet), may be cured at high 
temperature.
 This may manifest itself as the decrease of the mass 
$m_{\eta'}$,\cite{anomaly}
 for example.  The deconfinement will be better reflected in the 
properties of heavy-quark systems such as J$/\psi$\cite{miyamura} than in light
 hadrons.
 In this talk, we emphasize the significance of the  sigma meson in 
 QCD\cite{physrep,suppl},
 and  propose several experiments to produce 
the $\sigma$ in nuclei \cite{TIT,rcnp94,suppl} 
for confirming the existence of the 
 elusive particle and examining  the possible change 
of properties  of the meson due to a partial restoration of chiral symmetry 
in baryonic medium. 

\begin{table}[tbh]
\begin{center}
\begin{tabular}{|c|c|c|c|}   \hline
 Physics & Phenomena & Observables & Expectations \\ \hline \hline
  & Decrease of  & 2$\gamma $, 2$\pi$  & A bump in the  \\
 \ \ \         & $m_{\sigma}$  & from $\sigma $ & low mass region \\
        \cline{2-4} 
 Chiral &  Decrease of & dileptons        & Shift or smearing   \\
 Restoration  &  $m_{\rho },\ m_{\omega }$ & from  $\rho, \ \omega$ & 
  of the peak  \\
        \cline{2-4} 
  \ \ \     &  Decrease of & $K^{+}K^{-}$ and dileptons  & Reduction of the
 $K^{+}K^{-}$  \\
            &  $m_{\phi }$ & from  $\phi $ & yield \ \  \ \   \\
\hline
$U_A(1)$-   & Decrease of $m_{\eta '}$  & $2\gamma $ from $\eta '$
  &  Decrease of the  invariant \\
Restoration & \ \  & \ \  & mass \ \ \  \\
 \ \    & Change of  the & The branching ratio of & Increase of the mixing 
  \\
 \ \  & mixing property & $\eta $ and $\eta '$ from $J/\psi$ & angle
$\theta _{\eta}$ \ \ \  \\
\hline
Decon- &  Decrease of $m_{J/\psi}$ & dileptons from $J/\psi$ & Decrease of the
 invariant \\ 
finemnet & \ \ \                   & \ \ \            & mass\\ 
 \hline
\end{tabular}

\vspace{.5cm}

{Table 1.\ \ {\small  Interesting observables and their expected behavior
 in relation with the chiral transition,the 
possible restoration
  of the $U_A(1)$-symmetry and partial deconfinement 
at finite temperature and/or density. See the review \cite{physrep}
 for the relevant references.}}
\end{center}
\end{table}

\section{The Sigma Meson}

 The sigma meson is the chiral partner of the pion for the 
$SU_L(2)\otimes SU_R(2)$ chiral symmetry in QCD: In the 
$(1/2, 1/2)$-representation, the sigma field $\sigma $ constitutes the 
 quartet together with the three pion fields.  The order parameter of the 
chiral transition of QCD is the scalar quark condensate $\la \bar q q\ra\sim
 \sigma $, and the vacuum is determined as the sate where the effective 
 potential ${\cal V}(\sigma)$ takes the minimum.  Let us denote the minimum 
point by $\sigma _0$. Then the   particle representing the quantum fluctuation
 $\tilde {\sigma}\sim \la :(\bar q q)^2:\ra$ is the sigma meson. ($\sigma
 = \sigma _0 + \tilde{\sigma}$). In this sense, the sigma meson is
 analogous
 to the Higgs particle in the Weinberg-Salam theory as has been emphasized by 
the present author\cite{kek,physrep,suppl}.

 In  the ladder QCD\cite{scadron} and the Nambu-Jona-Lasinio model 
\cite{njl} 
 as a low-energy  effective theory\cite{physrep}, the sigma meson has the  
mass $m_{\sigma}$  almost twice of the constituent quark mass $\sim $ 350 MeV, 
 hence $m_{\sigma}\sim 700$ MeV. 
  Such a  scalar meson with a low mass can account for various hadron 
 phenomena\cite{suppl} including the 
 $\Delta I= 1/2$ rule for 
 the decay process K$^{0} \rightarrow \pi^{+}\pi^{-}$ or $\pi^{0}\pi^{0}$ 
\cite{morozumi}, the state-independent attraction in the 
intermediate range in  nuclear force\cite{durso}, 
 quark contents of the nucleon in the scalar channel and the $\pi$-N
 sigma term $\Sigma _{\pi N}$\cite{physrep} and so on; a nice summary is
 given in  \cite{suppl} for 
 the significance of the sigma meson in nuclear and hadron physics.

Experimentally, there is, however,  a controversy on the identification of the
  nonet scalar mesons
 in the particle zoo, and some people are skeptical even about the existence 
of the 
sigma  meson with a rather low mass, say about 600 to 800 MeV. 
Such skepticism may be attributed to the facts that the decay of the sigma 
 meson to two pions gives 
 rise to a huge width $\Gamma= 400 \sim 1000$ MeV of the sigma meson, and   
 that a possible coupling with glue balls with $J^{PC}= 0^{++}$ make the 
 situation obscure.  Therefore it is remarkable that recent extensive 
works
 on the phase shift analysis of the $\pi$-$\pi$ 
 scattering in the scalar channel
 claims a pole of  the scattering matrix in the complex energy plane  
 with the real part Re\ $m_{\sigma}=500-700$ MeV and the imaginary part
 Im$m_{\sigma}\simeq 500-800$MeV.\cite{kamin}
 There is  also preliminary experimental result at KEK\cite{shimizu},
 which seems to show a
 bump around 600 MeV with a width $\sim 400$ MeV in the reaction 
$\pi^{-}$p $\rightarrow $ n$\pi^{0}\pi^{0}$.  The 2 $\pi ^{0}$ are detected by
 4 $\gamma$'s.  This is a clever experiment
 in the sense that  by confining to the 2$\pi^{0}$ channel, 
 one can reject the iso-vector
 channel where we would have a huge yield from the rho meson.

\section{The Sigma Meson  at $T\not=0$ and/or $\rho_B\not=0$}
 Effective theories of QCD\cite{physrep} show that 
 the sigma meson mass $m_{\sigma}$ decreases in association with the chiral 
 restoration in hot and/or dense 
medium, while the pion mass keeps its value in free space as long
 as the system is in the Nambu-Goldstone phase.
 Then the width of the $\sigma$ is also expected to decrease  due to the
 depletion of the phase space for the decay $\sigma \rightarrow 2\pi$.
  Thus one can 
 expect a better chance to see the sigma meson in a clearer way in hot and/or
 dense medium  than in the vacuum. Actually, the decrease of the mass is
 already considerable even in the nuclear density $\rho_0$,
 as shown in \cite{bernard}.
 

It is worth mentioning that  the 
Walecka  model\cite{saito} also predicts the decrease of the masses of 
the scalar meson 
 as well as the $\omega$ meson in hot and/or dense nuclear matter:  
 Saito, Maruyama and Soutome (SMS)\cite{saito} showed that the extent of 
 the decrease of 
 $m_{\sigma}$ at $\rho_0$ reaches as large as 20\% of that in the free space, 
  even at zero temperature. It should be mentioned here that there are
  attempts to ``derive''
the Walecka model in a context of chiral symmetry,
 although the original  Walecka model is not
 constructed to have chiral symmetry; see \cite{brown} and 
 references cited therein.  SMS showed that  it is crucial to
 include the effects of the
 Dirac sea of nucleons for obtaining the decrease of the meson masses. 
 This is in accordance with the fact that the decrease of $m_{\sigma}$
 is associated with a change in the vacuum structure, 
 i.e., the chiral restoration,  in the chiral models.
As is well known, there arises a $\sigma$-$\omega$ mixing for
 finite three momentum ${\bf q}$ in nuclear matter, which effect is included in
 SMS's calculation\cite{saito}.
  It should be emphasized that the scalar-vector 
mixing with finite density is a general feature not restricted to the Walecka
 model\cite{sv}. 
This enables us to create our sigma meson by photons in 
nuclei\cite{TIT,suppl}; see below. SMS calculated
 the dispersion relations $\omega_{\alpha}=\omega_{\alpha}(q)$ \ 
$(\alpha= \sigma, \omega)$ for the sigma and omega mesons in nuclear matter;
 see Fig. 1.
 A remarkable point is that $\omega_{\sigma}(q)$ is almost overlap with the 
 photon line $\omega_{\gamma} =q$ between $q_1\simeq 200$ MeV$/c$ and 
$q_2\simeq 600$ MeV$/c$ where $\omega_{\sigma}$ cuts the photon line, i.e., 
 $\omega_{\sigma}=q$.  This means that real
 photons  also can  create the $\sigma$ in nuclei provided that the scalar
 meson in the Walecka model can be identified with the $\sigma$ meson 
 relevant with chiral symmetry.

\begin{flushleft}
\begin{minipage}[t]{8cm}
\end{minipage}
\end{flushleft}
\begin{flushright}
\begin{minipage}[thb]{7cm}
\vspace{1cm}
{\cl {\bf Figure 1}}
{\small The solid lines show the dispersion relation of the sigma meson ($S$)
 $\omega_{\sigma}(q)$,
 the longitudinal($L$) and the transverse($T$) mode of the omega meson
 at the nuclear density $\rho_0$ at $T=0$; taken and 
 modified from a figure by Saito et al\cite{saito}. The photon line
 $\omega=q$ is also shown. The dashed lines show the dispersion relation
 at $T\not=0$.}
\vspace{1cm}
\end{minipage}
\end{flushright}
\section{Production Mechanism of the Sigma Meson}
As we have seen in the last section, the sigma meson may decrease the mass in 
 nuclei, and hence the width due to the depletion of the phase space volume
 for the decay $\sigma \rightarrow \pi\pi$.  
Thus we propose several types of experiment\cite{suppl}
 to produce the sigma meson in nuclei:
 one uses pions, another protons and light nuclei and the other electrons (and
 real photons).
 To detect the sigma, one may use 4 $\gamma$'s and/or two leptons. The latter
 process is  possible because a scalar particle can be converted to 
 a vector particle because of the scalar-vector mixing in the 
 system with   a finite baryonic density. This mixing  is well known in the 
 Walecka  model as mentioned above.
 Microscopically, the process is described by 
 $\sigma \rightarrow {\rm N} \bar {\rm N}({\rm p-h}) \rightarrow \gamma$,
 where p-h represents nucleon particle-hole excitations.
  Here $\sigma $ may be replaced by any scalar particle, and
 $\gamma$ any vector particle with the same quantum numbers other than spin
 and parity.

{\bf 1.\ A\ ($\pi$, 4$\gamma$\  N)\ A$'$}\\
 In this reaction, the charged pion ($\pi ^{\pm}$) is absorbed by a nucleon in
 the nucleus, then the nucleon emits the sigma meson, which decays into 
 two pions.  To make a veto for the two pions from the rho meson, the produced
 pions should be neutral ones which decay to four $\gamma$ 's.

{\bf 2. \ A\ (P, 4$\gamma$\  N)\ A$'$} \\ 
  The incident proton, 
 deuteron or $^3$He ... collides with a nucleon in the nucleus, then the
 incident particle will emit the sigma meson, which decays into two pions.
 One may detect 4 $\gamma$ 's from 2 $\pi ^0$.  The 
 collision  with a nucleon may occur after  the emission of the sigma meson;
 the  collision process is needed for energy-momentum matching.

In the detection, one may observe the two leptons from the process 
 $\sigma \rightarrow {\rm N} \bar {\rm N}({\rm p-h})$ $\rightarrow \gamma$
 mentioned above:
 This detection may gives a clean data, but the yield might be small.

 {\bf 3. \  A\ (e$^{-}$, 4$\gamma$\   e$^{-}$)\ A$'$} \\ 
 The final example uses the electron beam:
 The $\gamma$ ray emitted from the electron is  converted to the omega
 meson in accord with the vector meson dominance principle. The omega meson
 may decay into the sigma meson in the baryonic medium via the process
 $\omega \rightarrow $ N $\bar{\rm N}({\rm p-h}) \rightarrow \sigma$.
 The sigma will
 decay into two pions. One may detect the 4$ \gamma$'s from the 2 neutral 
 pions. As noted in the previous section, real photons may be used to 
create $\sigma$ meson for some kinematical region, say $q=400 - 700$ MeV$/c$.

\section{Brief Summary}
We have summarized some  phenomena and experiments  which are to be 
 pursued in the project of ``Quark Nuclear Physics''. 
 We have emphasized the importance  to 
catch the tail of the elusive sigma meson somehow.
 Although it might be hesitating
 to dare to perform an experiment for possible production of such an elusive
 particle,  one may have a good chance to see a clear 
 signal for the particle in the baryonic medium, because
the sigma meson  may become  rather sharp in nuclei
 in association with partial restoration 
 of the chiral symmetry.  We have proposed several experiments to produce the
 $\sigma$ meson in nuclei: Some of them is based on the large
  scalar-vector mixing at $\rho_B$ for $\vec{q}\not=0$.  The detection
 should be done by observing neutral pions ($2\pi^0$).

In conclusion, the author acknowledges  
 Hajime Shimizu for his interest in this work.
This work is partially  supported by the Japanese Grant-in-Aid for Science
 Research Fund of the Ministry of Education, Science and Culture, No. 07304065

\end{document}